\PassOptionsToPackage{svgnames,dvipsnames}{xcolor}
\documentclass[runningheads,a4paper]{llncs}

\usepackage{paralist}
\usepackage{subfigure}

\usepackage[utf8]{inputenc}
\usepackage{emerald}
\usepackage{xargs}
\usepackage{xspace}
\usepackage{url}

\usepackage{todonotes}
\usepackage[nomargin,multiuser,inline,draft]{fixme} 
\fxusetheme{color}

\usepackage{rotating}

\usepackage{cmll}                   \usepackage{colonequals}            \usepackage{environ}                \usepackage{etex,etoolbox}          \usepackage{graphicx}

\usepackage{multirow}               \usepackage{nicefrac}
\usepackage{proof}                  \usepackage{bussproofs}
\EnableBpAbbreviations

\usepackage{thmtools, thm-restate, thm-autoref}  \usepackage{xifthen}                \usepackage{appendix}               

\usepackage{stmaryrd}
\usepackage{pdfsync}

\PassOptionsToPackage{usenames,dvipsnames,svgnames,table}{xcolor}
\usepackage{xcolor} \usepackage{tikz}                   \usetikzlibrary{shadows,arrows,shapes,automata,positioning,decorations.markings}

\usepackage{amsmath,amssymb,amsbsy}
\usepackage{mathtools} 

\usepackage{hyperref}
\hypersetup{
  colorlinks = true,
  linkcolor = purple,
  urlcolor  = purple,
  citecolor = teal,
  anchorcolor = teal
}
\usepackage[capitalise]{cleveref}
\crefformat{enumi}{condition~#2#1#3}
\crefname{fact}{Fact}{Facts}
\Crefname{fact}{Fact}{Facts}
\crefformat{fact}{Fact~#2#1#3}

\DeclareMathSizes{10}{10}{6}{6}

\usepackage{xargs}
\usepackage[normalem]{ulem} 

\usepackage{xifthen}        \newcommand{\ifempty}[3]{\ifthenelse{\isempty{#1}}{#2}{#3}}

\DeclareGraphicsExtensions{.png,.PNG,.pdf,.PDF,.jpg,.mps,.jpeg,.jbig2,.jb2,.JPG,.JPEG,.JBIG2,.JB2
}

\newif\ifemi

\newcommandx{\preprint}[3][1=preprint,2=Springer]{
  \ifempty{#1}{}{
	 \ \\[1em]\noindent
	 \textbf{Disclaimer}
    The published version of this paper is~\cite{#1} (\copyright\ #2).
  }
}

\usepackage{fixme}
\fxusetheme{color}
\FXRegisterAuthor{eM}{aeM}{\color{orange} {\underline{eM}}}

\usepackage[most]{tcolorbox}
\newtcolorbox{markbox}{
  enhanced,
  breakable,
  size=minimal,
  parbox=false,
  after={\par},
  before upper={\indent},
  colback=white,
  overlay = {
	 \draw[line width=2pt]
	 (frame.north east) -| ([xshift=3mm]frame.east) |-(frame.south east);
  },
  overlay first={\draw[line width=2pt] (frame.north east) -| ([xshift=3mm]frame.south east);},
  overlay middle={\draw[line width=2pt] ([xshift=3mm]frame.north east) -- ([xshift=3mm]frame.south east);},
  overlay last={\draw[line width=2pt] ([xshift=3mm]frame.north east) |- (frame.south east);},
}

\usepackage{todonotes}
\newcommand{\eMcomm}[2][check]{\ifthenelse{\equal{#1}{new}}{{\color{red}#2}}{\ifthenelse{\equal{#1}{changed}}{{\color{teal}{#2}}}{\ifthenelse{\equal{#1}{rm}}{\todo[color=black!3]{\tiny eM: removed\\'{#2}'}}{\todo[color=orange!20]{\tiny eM: \color{NavyBlue}#1}{\color{OliveGreen}{#2}}}}}}

\newcommand{\hidden}[1]{}
\newcommand{\hide}[1]{}

\newcommand{\cf}[2]{
  \fontsize{#1}{#1}{\selectfont{#2}}
}

\makeatletter
\newcommand{\dolist}[2]{\def\nextitem{\def\nextitem{#1}}\@for \el:=#2\do{\nextitem\textbf{\el}}}

\newcommand{\domathlist}[2]{\def\nextitem{\def\nextitem{\ensuremath{#1}}}\@for \el:=#2\do{\ensuremath{\nextitem}\textbf{\el}}}

\def\mktest#1{
  \def\transform##1+##2+##3{##1 piu' ##2 * ##3\penalty0}\do{\expandafter\transform#1}}
\def\mksubscript#1{
  \def\transform##1[##2]{##1_{##2}\penalty0}\do{\expandafter\transform#1}}
\newcommand{\mapcmd}[3][{, }]{\def\nextitem{\def\nextitem{#1}}\@for \el:=#3\do{\nextitem{#2{\el}}}}
\makeatother

\ifemi
\usepackage{showlabels}

\newcommand{\emi}[2]{
  \marginpar{\fcolorbox{red}{shadecolor}{\cf{#1}{{#2}}}}
}
\newcommand{\emic}[2]{\par
  \fcolorbox{red}{shadecolor}{\parbox{\linewidth}{ 
      \color{gray}
      \begin{description}
      \item[{\color{blue} #2}]{\sf #1}
      \end{description}}}
}
\else
\newcommand{\emi}[2]{}
\newcommand{\emic}[2]{}{}
\fi

\newcommand{\noarg}{}
\newcommand{\mkfun}[4][\colorFun]{
  \newcommand{#2}[1][#4]{{#1\ensuremath{\mathsf{#3}}}\ifempty{##1}{\noarg}{({##1})}}}

\mkfun{\head}{hd}{}
\mkfun{\tail}{tl}{}

\newcommand{\conf}[1]{\ensuremath{\langle {#1} \rangle}}

{\bfseries}{\rmfamily}
{\bfseries}{\rmfamily}

\makeatletter
\newcommand*{\da@rightarrow}{\mathchar"0\hexnumber@\symAMSa 4B }
\newcommand*{\da@leftarrow}{\mathchar"0\hexnumber@\symAMSa 4C }
\newcommand*{\xdashrightarrow}[2][]{\mathrel{\mathpalette{\da@xarrow{#1}{#2}{}\da@rightarrow{\,}{}}{}}}
\newcommand{\xdashleftarrow}[2][]{\mathrel{\mathpalette{\da@xarrow{#1}{#2}\da@leftarrow{}{}{\,}}{}}}
\newcommand*{\da@xarrow}[7]{\sbox0{$\ifx#7\scriptstyle\scriptscriptstyle\else\scriptstyle\fi#5#1#6\m@th$}\sbox2{$\ifx#7\scriptstyle\scriptscriptstyle\else\scriptstyle\fi#5#2#6\m@th$}\sbox4{$#7\dabar@\m@th$}\dimen@=\wd0 \ifdim\wd2 >\dimen@
    \dimen@=\wd2 \fi
  \count@=2 \def\da@bars{\dabar@\dabar@}\@whiledim\count@\wd4<\dimen@\do{\advance\count@\@ne
    \expandafter\def\expandafter\da@bars\expandafter{\da@bars
      \dabar@ 
    }}\mathrel{#3}\mathrel{\mathop{\da@bars}\limits
    \ifx\\#1\\\else
      _{\copy0}\fi
    \ifx\\#2\\\else
      ^{\copy2}\fi
  }\mathrel{#4}}
\makeatother

\newcommand{\quo}[1]{\lq\lq {#1}\rq\rq}
\def\finex{{\unskip\nobreak\hfil
\penalty50\hskip1em\null\nobreak\hfil$\diamond$
\parfillskip=0pt\finalhyphendemerits=0\endgraf}}

\definecolor{shadecolor}{rgb}{1,0.99,0.9}
\definecolor{bg}{rgb}{0.95,0.95,0.95}

\newcommand{\abcattr}[1][a]{\textsf{#1}}
\newcommand{\abccond}[1][\rho]{#1}

\def\colorExp{\color{NavyBlue}}

\newcommand{\abcexp}[1][e]{\colorExp #1}
\newcommandx{\abctuple}[1][1 = t]{\llparenthesis{#1}\rrparenthesis}
\newcommandx{\abcget}[2][1=a,2={id},usedefault=@]{
  \ptp[{#1}]{\colorOp .}\abcattr[{#2}]
}
\newcommandx{\abcptp}[2][1=a,2=\abccond,usedefault=@]{\ifempty{#1}{}{\ptp[{#1}] \ifempty{#2}{}{{\colorOp \shortmid}}} {#2}}
\newcommandx{\abcint}[6][1=a,2=\abccond,3=e,4=e',5=b,6=\abccond',usedefault=@]{
  \abcptp[{#1}][{#2}]
  \ {\colorOp \xrightarrow{\scriptstyle \abcexp[#3]\quad\abcexp[#4]}}\ 
  \abcptp[{#5}][{#6}]
}
\newcommandx{\mkabcint}[8][3=a,4=\abccond,5=e,6=e',7=b,8=\abccond',usedefault=@]{
  \node[bblock, #1] (#2) {$\abcint[{#3}][{#4}][{#5}][{#6}][{#7}][{#8}]$};
}

\tikzset{
    abccallout/.style={
      fill=green!10,
		opacity=.5,
		overlay,
		align=center,
      cloud callout,
		cloud puffs=15,
		aspect=2.5,
		cloud ignores aspect,
		cloud puff arc=100,
		shading=ball
    }
  }

\newcommandx{\abcP}[6][1=P,2=K,3=.1cm,4=1cm,5=north east,6=proc,usedefault=@]{
  \begin{tikzpicture}
	 \node[fill=blue!10, shape=circle] (#6) {$\p[#1]$};
	 \node[abccallout, above = #3 of #6, xshift=#4, callout absolute pointer={(#6.#5)}] {$#2$}
	 ;	 
	 \draw[decorate,decoration={expanding waves,angle=7,segment length = .05cm}] (#6.east) -- ++(.5cm,0)
	 ;
  \end{tikzpicture}
}

\ExplSyntaxOn
\NewDocumentCommand{\ucgreek}{m}
 {
  \str_case:nn { #1 }
   {
    {A}{\mathrm{A}}
    {B}{\mathrm{B}}
    {C}{\Sigma}
    {D}{\Delta}
    {E}{\mathrm{E}}
    {F}{\Phi}
    {G}{\Gamma}
    {H}{\mathrm{H}}
    {I}{\mathrm{I}}
    {J}{\Theta}
    {K}{\mathrm{K}}
    {L}{\Lambda}
    {M}{\mathrm{M}}
    {N}{\mathrm{N}}
    {O}{\mathrm{O}}
    {P}{\Pi}
    {Q}{\mathrm{X}}
    {R}{\mathrm{P}}
    {S}{\Sigma}
    {T}{\mathrm{T}}
    {U}{\Upsilon}
{W}{\Omega}
    {X}{\Xi}
    {Y}{\Psi}
    {Z}{\mathrm{Z}}
   }
 }
\NewDocumentCommand{\lcgreek}{m}
 {
  \str_case:nn { #1 }
   {
    {a}{\alpha}
    {b}{\beta}
    {c}{\varsigma}
    {d}{\delta}
    {e}{\varepsilon}
    {f}{\varphi}
    {g}{\gamma}
    {h}{\eta}
    {i}{\iota}
    {j}{\vartheta}
    {k}{\kappa}
    {l}{\lambda}
    {m}{\mu}
    {n}{\nu}
    {o}{o}
    {p}{\pi}
    {q}{\chi}
    {r}{\rho}
    {s}{\sigma}
    {t}{\tau}
    {u}{\upsilon}
{w}{\omega}
    {x}{\xi}
    {y}{\psi}
    {z}{\zeta}
   }
 }
\ExplSyntaxOff

\newcommand{\toolidcol}{teal}\newcommand{\toolid}[1]{\textcolor{\toolidcol}{#1}}

\newcommand{\chorgramsite}{\url{https://bitbucket.org/eMgssi/chorgram/src/master/}}
\newcommand{\chorgram}{\textsf{\toolid{ChorGram}}\xspace}

\newcommand\SEArch{{\ECFAugie \toolid{SEArch}}\xspace}

\newcommand{\orcidlink}[1]{\href{https://orcid.org/#1}{\textcolor[HTML]{A6CE39}{\orcidID{#1}}}}

\usepackage{tikz}
\newcommand*\circled[1]{
     \tikz[baseline=(char.base)]{
          \node[shape=circle,draw,inner sep=0.5pt,fill=black] (char) {{\color{white} #1}};
     }
}

\title{
 \SEArch: an execution infrastructure for service-based software systems\thanks{The authors want to thank Ignacio Vissani for his indispensable contributions to the design of \SEArch.}
}
\titlerunning{\SEArch: an execution infrastructure for service-based software systems}
\author{Carlos G. Lopez Pombo 
	\thanks{On leave from Instituto de Ciencias de la computaci\'on CONICET--UBA and Departamento de Computación, Facultad de Ciencias Exactas y Naturales, Universidad de Buenos Aires.}\inst{1,2}\orcidlink{0000-0002-0248-5019} \and Pablo Montepagano\inst{3} \and Emilio Tuosto\inst{4}\orcidlink{0000-0002-7032-3281}
}
\institute{Centro Interdisciplinario de Telecomunicaciones, Electrónica,
  Computación y Ciencia Aplicada - CITECCA, Universidad Nacional de
  Río Negro - Sede Andina. \and 
  Consejo Nacional de Investigaciones Cient\'{\i}ficas y T\'ecnicas - CONICET.\\\email{cglopezpombo@unrn.edu.ar}  \and 
  Departamento de Computaci\'on, Facultad de Ciencias Exactas y Naturales, Universidad de Buenos Aires.\\\email{pmontepagano@dc.uba.ar} \and
  Gran Sasso Science Institute\\\email{emilio.tuosto@gssi.it}
} 
 \authorrunning{Carlos G. Lopez Pombo, Pablo Montepagano, Emilio Tuosto}
\index{Lopez Pombo, Carlos G.}
\index{Montepagano, Pablo}
\index{Tuosto, Emilio}

\makeatletter
\begin{document}

\maketitle

\begin{abstract}
  The shift from monolithic applications to composition of distributed
  software initiated in the early twentieth, is based on the vision of
  \emph{software-as-service}.
This vision, found in many technologies such as RESTful APIs,
  advocates globally available services cooperating through an
  infrastructure providing (access to) distributed computational
  resources.
Choreographies can support this vision by abstracting away local
  computation and rendering interoperability with message-passing:
  cooperation is achieved by sending and receiving messages.
Following this choreographic paradigm, we develop \SEArch, after
  {\ECFAugie S}ervice {\ECFAugie E}xecution {\ECFAugie Arch}itecture,
  a language-independent execution infrastructure capable of
  performing transparent dynamic reconfiguration of software
  artefacts.
Choreographic mechanisms are used in \SEArch to specify
  interoperability contracts, thus providing the support needed for
  automatic discovery and binding of services at runtime.
\end{abstract}

\section{Introduction}
\label{sec:introduction}

In the past two decades the paradigm generally known as
\emph{Service-Oriented Computing} (SOC) has become predominant in
software development.
This paradigm comprises many variants such as ---among others--- cloud
computing, fog and edge computing, and many forms of distributed
computing associated with what is know as the \emph{Internet of
  Things}.\footnote{Although the terminology has evolved, SOC is still widely used to
  denote these software systems.}
Key to service-oriented computing is the possibility of dynamically
search and combine distributed computational resources exposed as
services interacting over an existing communication infrastructure.
This vision of software systems can be found in applied technologies,
such as RESTful APIs, and has put forward what is commonly know as the
API Economy.
In this context many aspects of software development are facilitated,
but service integration becomes non-trivial.
A common interaction mechanism in SOC is (some form of remote) procedure
call (for instance using JSON-RPC or via the HTTP protocol)
since it resembles the typical function or procedure call of
programming languages.

Choreographic approaches~\cite{w3c:wsdl20} propose an alternative
interaction mechanism by conceptually separating the local
computations of the components from their communication aspects.
Under this approach, interoperability is understood at a more abstract
level decoupled from any computational aspect. Within choreographic
approaches we found the subclass of message-passing systems, a type of
system where cooperation is achieved by the simple actions of sending
and receiving messages through existing communication channels.

\emph{Asynchronous Relational Networks}~(ARNs)~\cite{fiadeiro:tcs-503}
yield a formalisation of the elements of an interface theory for
service-oriented software architectures.
More precisely, ARNs are a formal orchestration model based on
hypergraphs whose hyperedges are interpreted either as processes or as
communication channels.
The nodes (or points) that are only adjacent to process hyperedges are
called \emph{provides-points}, while those adjacent only to
communication hyperedges are called \emph{requires-points}: the former
constitute the interface through which a service exports its
functionality while the latter yields the interface through which an
activity expects certain service to provide a functionality.
In the operational semantics of ARNs given in~\cite{vissani:wadt14-f}
actions performed by a component can dynamically trigger an automatic
and transparent process of discovery and binding of a compliant
service.
The composition of ARNs (i.e., how binding is viewed from a formal
perspective) is obtained by \quo{fusing} provides-points and
requires-points, subject to a certain compliance check between the
contract associated to them.
Later, \cite{vissani:places15} used \emph{communicating finite state
  machines} (CFSM)~\cite{brand:jacm-30_2} as a formal language for
determining service interoperability automatically.

More recently, \cite{senarruzzaanabia:mathesis} has proposed
data-aware CFSMs, an extension of CFSMs with \emph{assertions}, namely
first-order formulae associated to the communication actions.
Besides, \cite{senarruzzaanabia:mathesis} has introduced a
bisimulation relation for data-aware CFSMs and implemented an
algorithm for checking bisimilarity of data-aware CFSMs.
In this setting, given participants {\tt A} and {\tt B}, and a first
order formula $\alpha(x)$, where $x$ is a free variable, an action
${\tt AB}!y\conf v\ |\ \alpha (x)$ is interpreted as: participant {\tt
  A} \underline{sends to} participant {\tt B} a message of type $y$
with value $v$ \underline{guaranteeing} that $\alpha (v)$ holds.
Dually, ${\tt AB}?y\conf v\ |\ \alpha (x)$ is interpreted as
participant {\tt A} \underline{receives from} participant {\tt B} a
message of type $y$ with value $v$ \underline{assuming} that
$\alpha (v)$ holds.
The rationale behind data-aware CFSMs is that assertions act as
functional contracts predicating over the data exchanged by the
components that participate in the communication.

In this work we introduce the language-independent execution
infrastructure \SEArch, after {\ECFAugie S}ervice {\ECFAugie
  E}xecution {\ECFAugie Arch}itecture, based on the operational
semantics given to ARNs and the interoperability and functional
compliance criterion supported by data-aware CFSMs.
In particular, we give the architecture of \SEArch
(\cref{sec:conceptual}) and its main implementation details
(\cref{sec:implementation}).
Also, we showcase \SEArch on an on-line business cart
(\cref{sec:casestudy}).
In~\ref{sec:conclusions} we draw some conclusions and discuss further
lanes of research.

 \section{A conceptual view of \SEArch}
\label{sec:conceptual}
There is a wide range of service-oriented
architectures (SOAs) dictating design principles for SOC, each one
with its own
idiosyncrasy~\cite{mulesoft:soa,ibm:soa,microsoft:soa,oracle:soa}.
We embrace those that hinge on three main concepts: a \emph{service
  provider}, a \emph{service client}, and a \emph{service broker}.
The latter handles a \emph{service repository}, a catalogue of service
descriptions searched for in order to discover services required at
runtime.
In fact, the service broker is instrumental to the \emph{discovery} of
services according to a contract and of their \emph{binding}, the
composition mechanism that permits to \quo{glue} services together at
runtime as advocated by some SOAs.

To support SOC, \SEArch offers a mechanism for populating registries
and composing service-based application.
Registering a service is, in principle, very simple: the service
provider sends the service broker a request for registering a service
attaching a (signed) package containing the contract and the unique
resource identifier (URI) of the provided service.

\begin{figure}[t]
  \centering
  \includegraphics[width=0.7\textwidth]{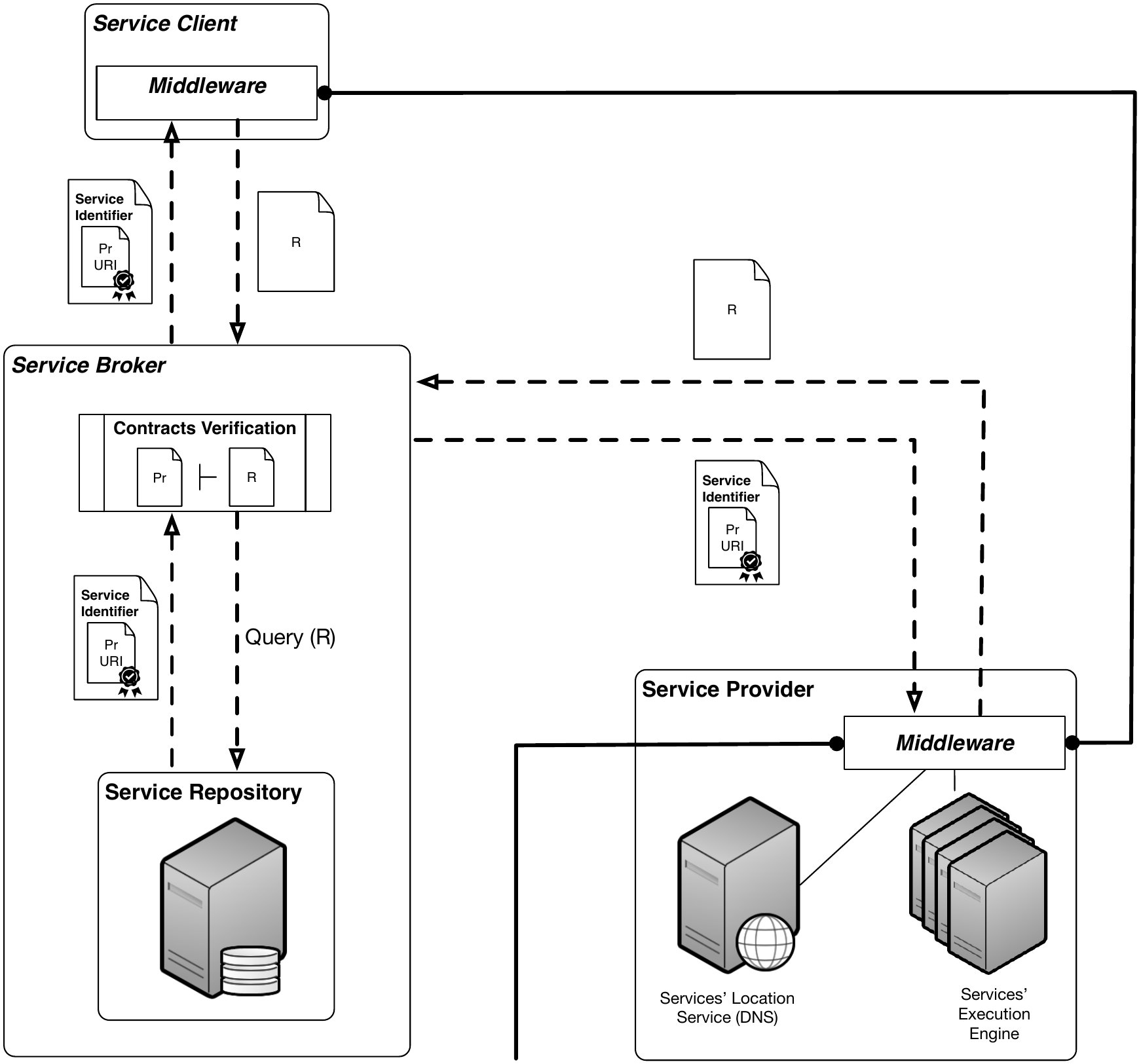}
  \caption{Service execution procedure in \SEArch}
  \label{fig:execution}
\end{figure}

The execution process of a service-based system in \SEArch is
significantly more complex.
\Cref{fig:execution} depicts the workflow.
When launched, a component registers their communication channels to
its \emph{middleware}, each of which has its corresponding contract
formalised as a set of data-aware CFSMs, one for each requires-point.
This is required because the middleware has to mediate the
communication with other components.
In fact, when the component a running application, say $C$, tries to
interact with another component, the middleware $C$, say $M$, captures
the attempt and checks whether the communication session for that
communication channel has been created.
If no such session exists, the dynamic reconfiguration process
triggers as follows:
\begin{enumerate}
\item $M$ sends the service broker the contract of the
  communication channel;
\item for each data-aware CFSM $\mathtt{R}$ in the contract, the
  service broker queries the service repository for candidates;
\item the service repository returns a list candidates in the form of
  $\langle \mathtt{Pr}, u \rangle$, where $\mathit{Pr}$ is a
  data-aware CFSM and $u$ is the URI of the service\footnote{\SEArch
	 is parametric in the implementation of the service repository so
	 we assume it is not capable of checking compliance using
	 behavioural contracts. We only relay on its capability of
	 returning a list of candidates, obtained by using potentially more
	 efficient and less precise criteria, for example, an ontology.};
\item the service broker checks whether the provision contract
  $\mathtt{Pr}$ is bisimilar to the requirement contract $\mathtt{R}$;
\item once the service broker has found services satisfying all the
  requirement contracts, it returns the set of URIs to $M$;
\item $M$ opens a communication with the service middleware of each
  service returned by the provider requiring the execution of an
  instance of the corresponding service.
\end{enumerate}
Then, $M$ sends to or receives from the service middleware of the
partner component the actual message; and the execution process
proceeds.
Notice that the execution of the service might also have its own
requirements putting it as the originator a new dynamic
reconfiguration.

The schematic view discussed before establish several requisites over
the implementation of middleware and the service broker.
We organise the discussion by considering these elements and their
role in the execution architecture:

\paragraph{{\bf The middleware}} provides a private and a public
interfaces.
The former implements functionalities accessible by service clients
and service providers.
The public interface implements the capabilities needed for
interacting with service brokers and other service middlewares.

The private interface consists of:
\begin{itemize}
\item \texttt{RegisterApp} to register a service and expose it in the
  execution infrastructure. This functionality opens a bidirectional
  (low level) communication channel with the service middleware that
  will remain open in order to support the (high level) communication
  with other services.
\item \texttt{RegisterChannel} to register communication channels expressing requirements.
This functionality  provides the middleware with the relevant information for triggering
  the reconfiguration of the system and managing the communications.
The functionality can be used by any software artefact running in
  the host, regardless if it a service or client application.
\item \texttt{AppSend} / \texttt{AppRecv} to communicate with partner
  components
\item \texttt{CloseChannel} to close a communication channel.
\end{itemize}
The reader should note the asymmetry between the existence of a
function for explicitly closing a communication sessions, and the lack
of one for opening it.
The reason for this asymmetry resides in that, on the one hand,
transparency in the dynamic reconfiguration of the system is a key
feature of \SEArch but, on the other hand, it is in general not
possible to determine whether a communication session will be used in
the future.

The public interface consists of:
\begin{itemize}
\item \texttt{InitChannel} to accept the initiation of a
  point-to-point (low-level) communication channel.
This operation allows the service broker to initiate the
  communication infrastructure that will connect the service executing
  behind the service middleware to the other participants in a
  communication session being setup.
\item \texttt{StartChannel} to receive notification about point-to-point
  communication channels.
This operation formally notifies the service middleware that the
  brokerage of participants according to a communication channel
  description was successful and the communication session has been
  properly setup.
\item \texttt{MessageExchange} to exchange messages between service
  middlewares.
\end{itemize}

\begin{figure}[ht]
  \centering
  \includegraphics[width=\textwidth]{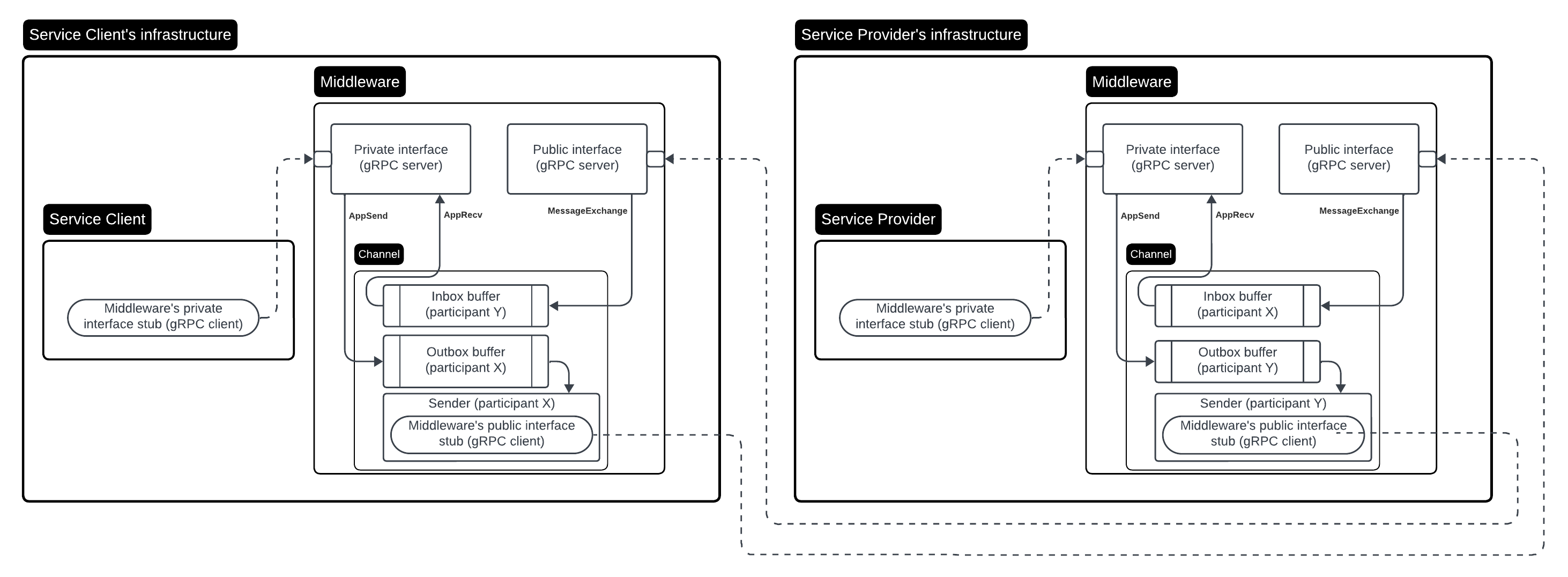}
  \caption{Structural design of the point-to-point communication between a service client and a service provider.}
  \label{fig:point-to-point-communication}
\end{figure}

\Cref{fig:point-to-point-communication} shows the application infrastructure and how buffers are used to provide
point-to-point communication with external services. Within the infrastructure, it is possible to identify the structural design of the middleware.

\paragraph{{\bf The service broker}}
exposes only two functionalities in a public interface:
\begin{itemize}
\item \texttt{BrokerChannel} to issue requests for brokerage.
This operation allows a service middleware to request for the
  brokerage of communication channel, and the subsequent creation of a
  communication session over which the chosen services will
  communicate.
\item \texttt{RegisterProvider} to issue requests for the registration
  of a service provider.
This operation if the external counterpart of the functionality
  \texttt{RegisterApp} through which the service middleware provides
  the service providers the possibility of being offered as services
  available in the execution infrastructure.
\end{itemize}

\begin{figure}[ht]
  \centering
  \includegraphics[width=0.7\textwidth]{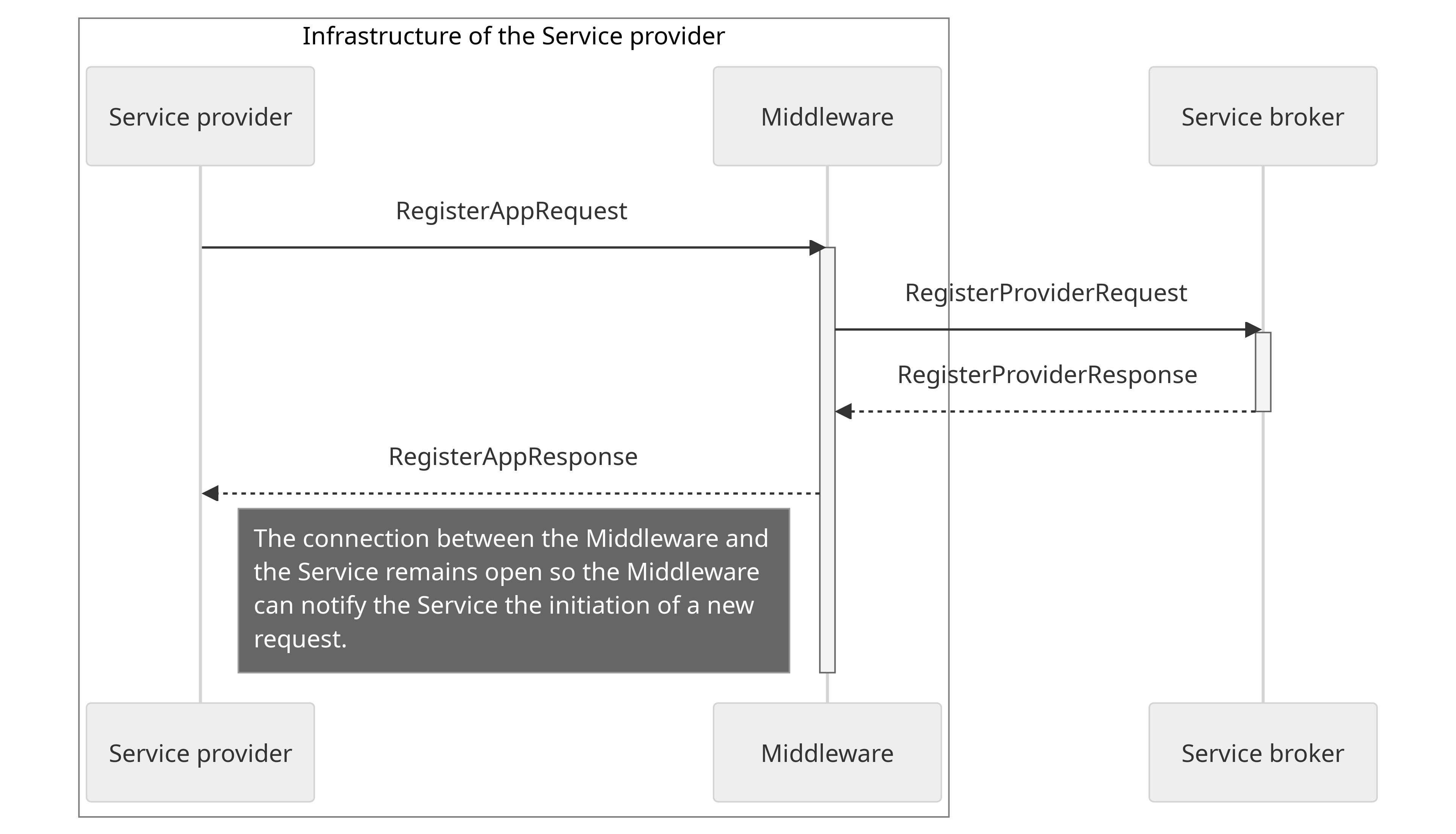}
  \caption{Sequence diagram of the process of registration of a service.}
  \label{fig:seq-diagram-registration}
\end{figure}

\Cref{fig:seq-diagram-registration} show the sequence diagrams offering a high level view of
the processes of registration of service to the service broker.

\begin{figure}[ht]
  \centering
  \includegraphics[width=\textwidth]{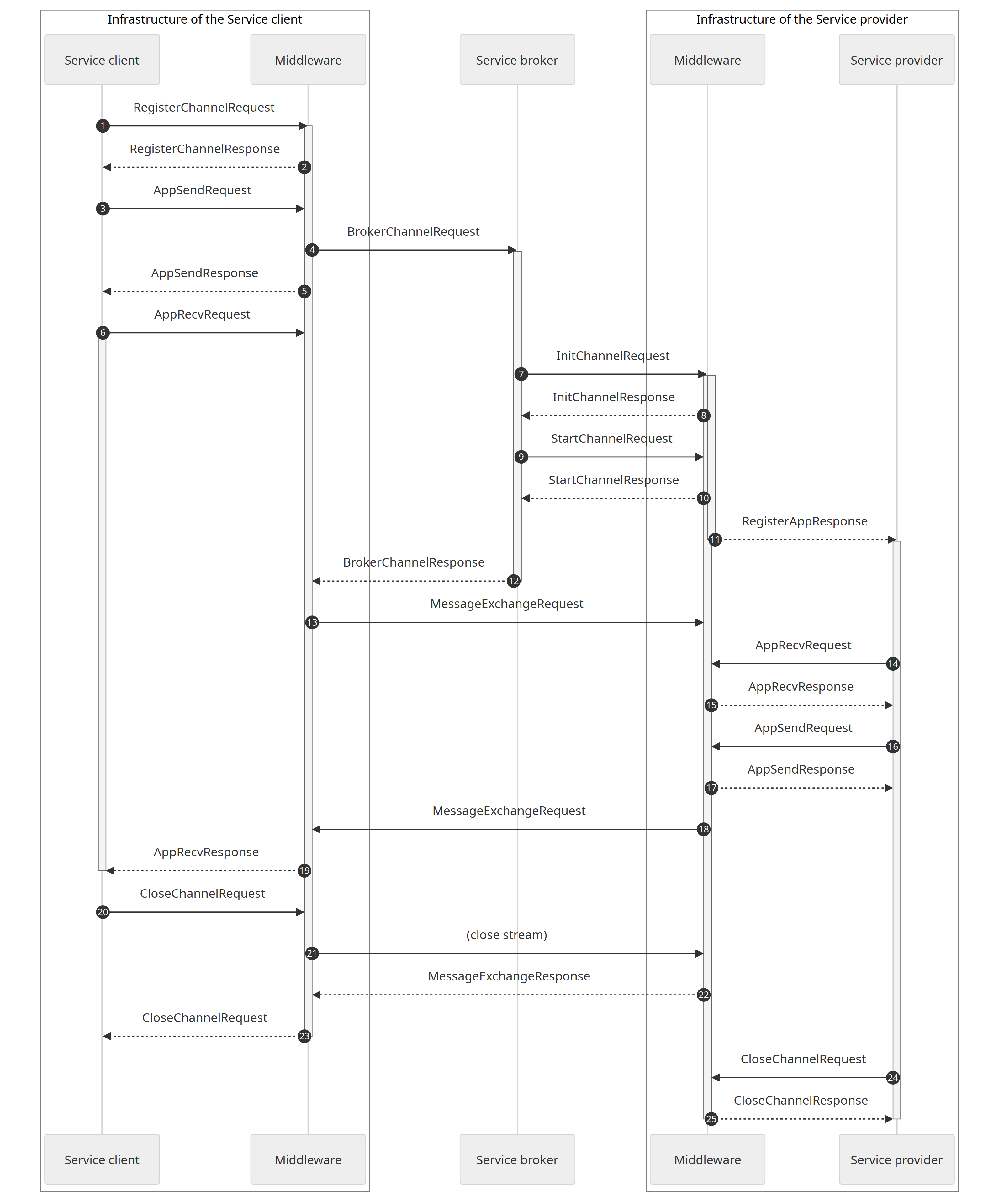}
  \caption{Sequence diagram of the process of brokerage of a communication channel.}
  \label{fig:seq-diagram-brokerage}
\end{figure}

\Cref{fig:seq-diagram-brokerage} shows the process of brokerage of a
communication channel given the interfaces of the middleware
and the service broker, detailed above in this section.

The process of brokering a communication channel for building a
session (cf. \cref{fig:seq-diagram-brokerage}) is significantly more
complex.
The service client uses the communication channel in the message
\texttt{AppSend} (step \circled{3}).
Concurrently, the service middleware begins the brokerage
process by sending the contract to the service broker 
(step \circled{4}) and
queues the message while acknowledging the service client with a
message of type \texttt{AppSendRespond} (step \circled{5}).
If the service client has to receive a message (\texttt{AppRecv}), the
middleware captures the attempt triggering the brokerage and going
through the same process for initiating the communication session.
In this case, the service client will remain blocked until the
expected message arrives.

The service broker, upon receiving the contract, queries the service
repository for candidates and executes the compliance checks.
Each compliance check can be too costly so the service broker
implements a cache for storing precomputed positive results.

After choosing concrete providers for the participants in the contract, the service broker performs two successive rounds of messages with the chosen service provider.
In the first round, a message of type \texttt{InitChannelRequest} is
sent (step \circled{7}) to tell the service middlewares that a
communication session involving its service provider is being
initiated; the message also contains the URIs of all the other
participants in the communication session.
At the same time, this message allows the service broker to verify
that the provider is indeed online.
Upon reception of this message, a service middleware must accept
incoming messages for this channel and queue them for the eventual
reception by the service provider.
If all the service providers respond successfully with a message of
type \texttt{InitChannelResponse} (step \circled{8}), then the service
broker performs a second round with a messages of type
\texttt{StartChannelRequest} (step \circled{9}), to confirm that the
communication session has been initiated.

After receiving both initialization messages, the service middleware
sends the service provider a message of type
\texttt{RegisterAppResponse} (step \circled{{\scriptsize 11}}) containing the UUID
of the new communication session.
Then, each service provider can start communicating over this session
according to their contracts.
Once a session is initiated, the service middlewares establish
unidirectional streams with each other to send messages.
In \cref{fig:seq-diagram-brokerage} the service middleware of the
service client opens a stream with the other service middleware by
sending a message of type \texttt{MessageExchange} (step
\circled{{\scriptsize 13}}).
After the service provider has received the message (steps
\circled{{\scriptsize 14}}-\circled{{\scriptsize 15}}), it sends a message (steps
\circled{{\scriptsize 16}}-\circled{{\scriptsize 17}}) forcing the service middleware of the
service provider to establishes a stream in the opposite direction
(step \circled{{\scriptsize 18}}).

Finally, the service client can close the channel by sending a message
of type \texttt{CloseChannelRequest} (step \circled{{\scriptsize 20}}) to its
service middleware, closing the stream used to communicate with the
service middleware of the service provider.

 \section{Implementation}
\label{sec:implementation}

The main objective of \SEArch\footnote{Available at
  \url{https://github.com/pmontepagano/search}.} is to provide
transparent integration of services offering abstractions for the
interoperability of heterogeneous software artefacts.
This is mainly achieved by the middleware featured by \SEArch, which
mediates all the interactions between software components.
To this end, we implemented the lower layer of the communication infrastructure over gRPC\footnote{Available at \url{https://grpc.io}.}, Protocol
Buffers\footnote{Available at \url{https://protobuf.dev/}.}.
The former is a high-performance RPC framework offering an easy and
scalable solution to the problem of microservices integration; the
latter is a typed and structured data packet serialisation format,
used as an interface description language, which provide a high-level
solution for system level communication.\footnote{Although widely used,
  HTTP is not an ideal option for \SEArch due to two severe
  limitations: HTTP supports request-response and it has no native
  support for typed messages (schemas).}
Both gRPC and Protocol Buffers aim to provide a general, yet easy to
use, tool for developing communication infrastructure between
systems. For this reason, there are compilers that interpret message
and service definitions from Protocol Buffer \texttt{.proto} files, and generate code in a wide variety of
programming languages for manipulating those messages with native
classes and types.\footnote{Until July of 2023 gRPC and Protocol
  Buffers officially supports {C\#} / {.NET}, {C++}, {Dart}, {Go},
  {Java}, {Kotlin}, {Node}, {Objective-C}, {PHP}, {Python} and {Ruby}
  among others; \url{https://grpc.io/docs/languages/}.}

We implemented \SEArch in Go~\cite{donovan15}, a language with native support for channel-based concurrency (goroutines) and gRPC. Such flexibility is of particular relevance to us as it provides a high degree of portability, specially in order to cope with the heterogeneity of the computational resources available, that could be integrated to the \SEArch ecosystem. 

As said, we adopt CFSMs to model contracts for interoperability; more
specifically, we use an implementation in Go of
CFSMs\footnote{Available ar \url{https://github.com/nickng/cfsm}.} which we extended with a bisimilarity test.
Test for bisimilarity is used by \SEArch service brokers as interoperability compliance
criterion when selecting service providers.
There exists extensions of the CFSMs enabling their use for describing
functional aspects of participants~\cite{senarruzzaanabia:mathesis}
and quality-of-service non-functional aspect~\cite{lopezpombo:ictac23}, but they were not yet been
added to \texttt{nickng/cfsm} so, for the sake of this presentation,
the compliance check will only reflect interoperability.
Choosing Go as a programming language was key for building a solution
satisfying the most important hypothesis of the computational model;
the order of messages is preserved.
This hypothesis is vital to the correctness of a message-passing
communicating systems, thus it becomes an important constraint of the
implementation.
Channels in Go are similar to Unix pipes, being thread-safe FIFO
queues.
This, together with the use of an RPC of type stream, allowed for the
implementation of inbox and outbox buffers for each participant,
together with a sender routine in charge of processing the Outbox;
also providing a simple implementation for \texttt{MessageExchange}
and \texttt{AppRecv} which essentially act as enqueue and dequeue
operations, respectively.

\begin{figure}[ht]
  \centering
  \includegraphics[width=\textwidth]{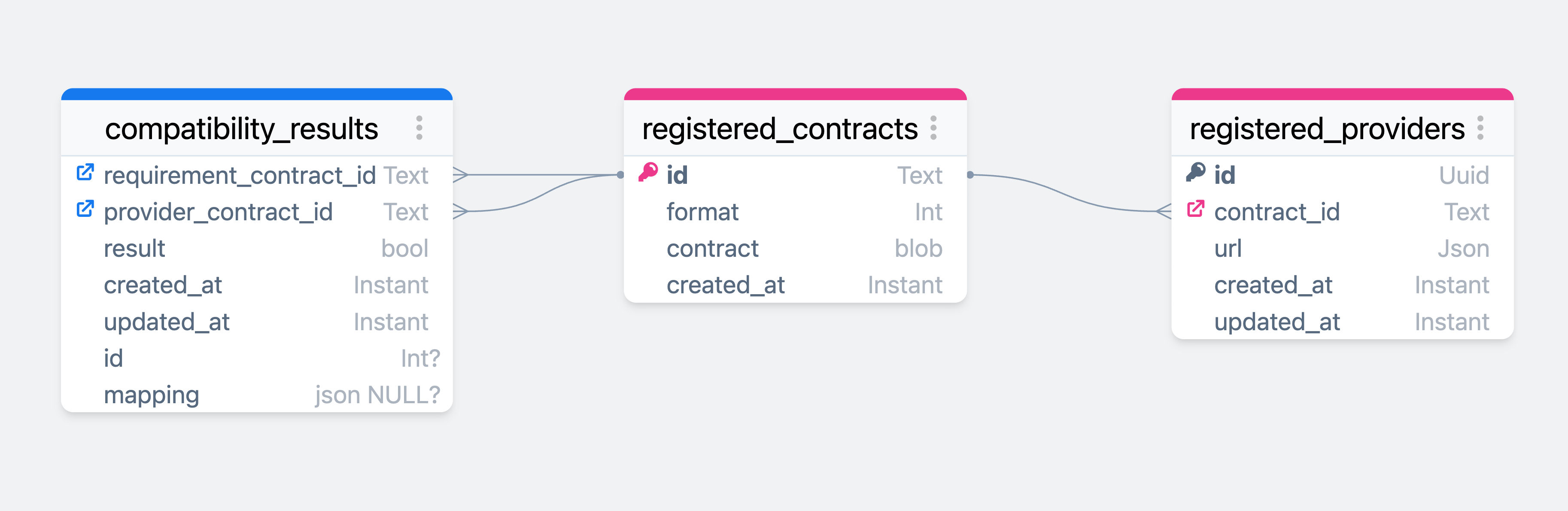}
  \caption{ORM schema definition for the cache.}
  \label{fig:er-cache}
\end{figure}

In general, testing bisimilarity of CFSMs is computationally costly,
resulting in a bottleneck. To tackle this problem the service broker
features a cache associating lists of compliant services to
requirement contract.
The implementation consists of a very simple schema shown in
\cref{fig:er-cache} done as an \emph{object-relational mapping}
(ORM) in \texttt{ent}\footnote{Available at \url{https://entgo.io}.},
a simple and powerful entity framework specifically designed for Go.
Whenever a requirement contract produces a miss in the cache table, or
when the service repository returns candidates that have not been
checked, the compliance checks are performed concurrently by launching
separate goroutines\footnote{The library \texttt{conc} (available at
  \url{https://github.com/sourcegraph/conc}) was used to prevent leaks
  (routines that execute indefinitely).}.
For the moment, the concurrent execution of compliance checks only profits from parallelism locally (multi-core and multi-CPU servers) but it provides no support for multi-server architectures like clusters, cloud computing, etc.

 \section{An online credit card payment service}
\label{sec:casestudy}

\begin{figure}[h]\centering
  \begin{tikzpicture}
\node[scale=.4] (c){\includegraphics{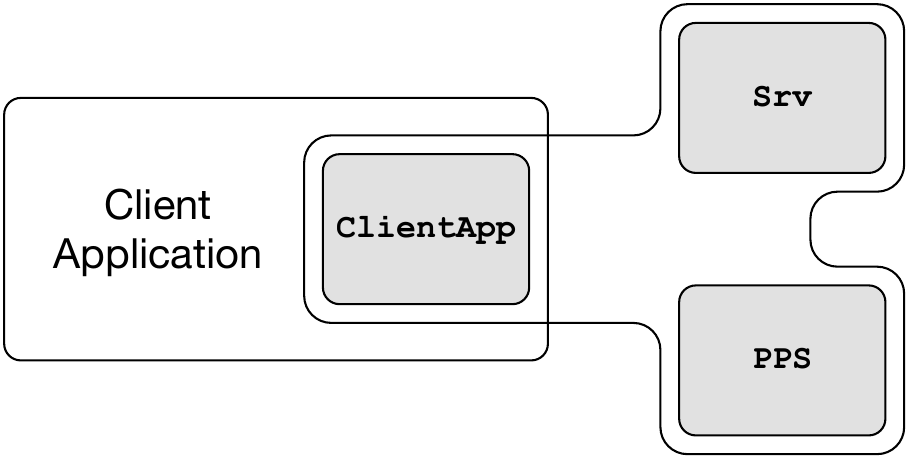}};
	 \node[right = of c, scale =.35,yshift=-2.6cm] (s){\includegraphics{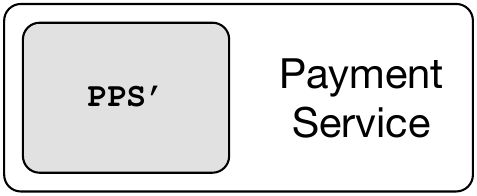}};
	 \node [above = .6cm of s, scale = .35]{\includegraphics{server}};
  \end{tikzpicture}
  \caption{A view of Client (left), seller (top right), and payment service (bottom right)}
  \label{fig:payment-system}
\end{figure}

This section shows a case study where an online shop relies on a third
party payment service.
The application involves three participants: a client (developed in
Java), the seller (a backend server developed in Python), and a
payment service (developed in Go).
\Cref{fig:payment-system} shows an abstract view of the three
components where contracts are represented by the gray boxes (The source
code of these components can be found at
\url{https://github.com/pmontepagano/search/examples/credit-card-payments}).
The rotated V-shape box represents the client's communication channel
specified by the contracts \texttt{ClientApp}, \texttt{Srv}, and
\texttt{PPS}, where the latter two are to be interpreted as
requirements to be fulfilled by other participants.
The contracts of seller and payment service are to be interpreted as
provisions by the corresponding services.

We now detail each component.

\subsection{Client application}
As said, the client component is implemented in Java.
We focus on how channel registration and communication is rendered in
the client.

\begin{figure}[h]\centering
 \includegraphics[width=0.9\textwidth]{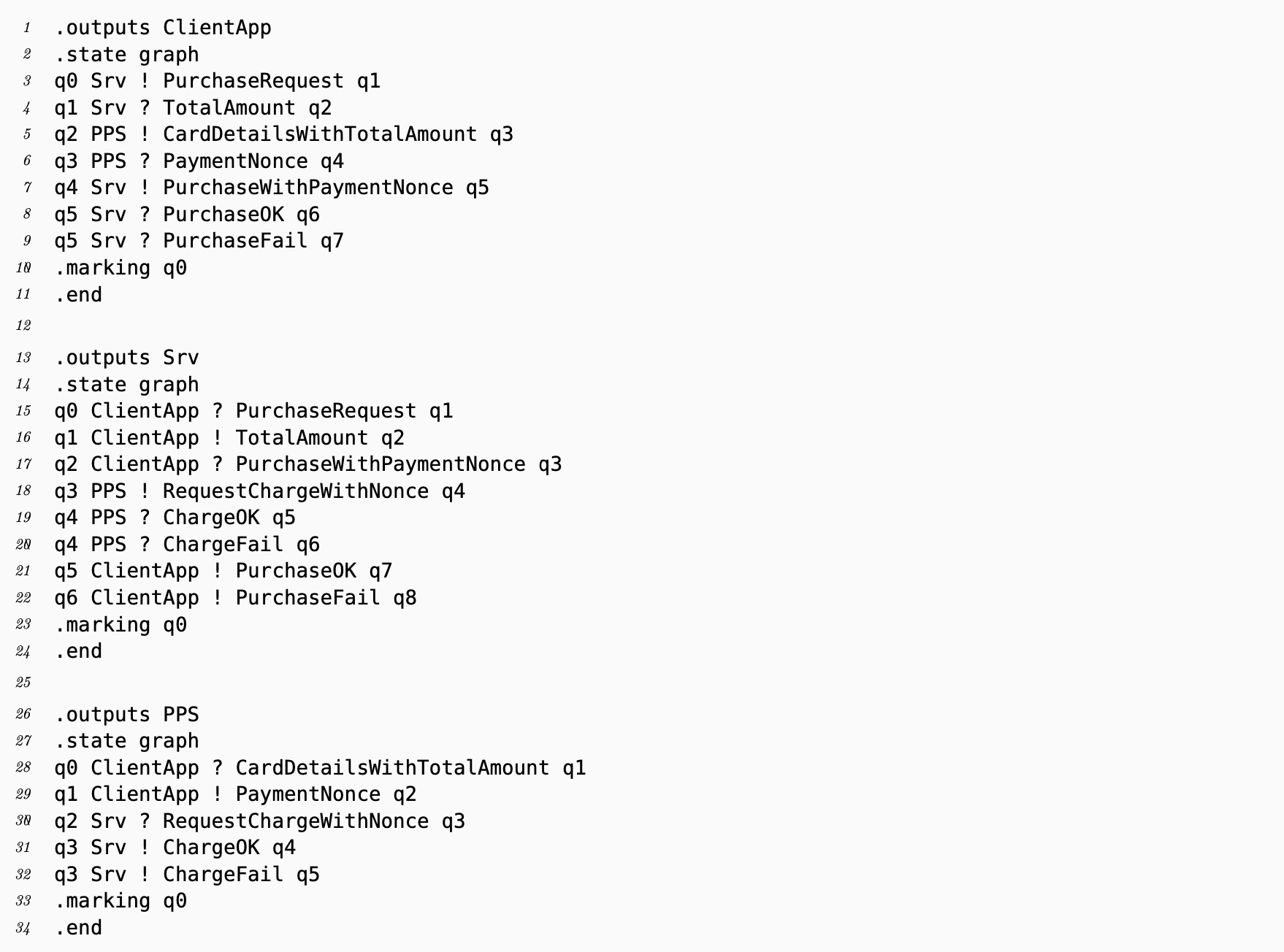}
 \caption{Communication channel specification}
 \label{fig:channel-spec}
\end{figure}

The communication channel of the client consists of the CFSMs in
\cref{fig:channel-spec}.
This specification dictates that \texttt{ClientApp} starts by sending
a \texttt{PurchaseRequest} to the seller (\texttt{Srv}) and, after
receiving the \texttt{TotalAmount} from the seller, the client sends
\texttt{CardDetailsWithTotalAmount} to the payment server
(\texttt{PPS}); then the purchase is completed as shown in
\cref{fig:channel-spec}.
The other CFSMs behave accordingly.

Below we report the snippet for the creation of the communication
channel between the client application and its middleware's private
interface by resorting the Java stubs generated by Protocol
  Buffer.
\begin{figure}[h]\centering
 \includegraphics[width=\textwidth]{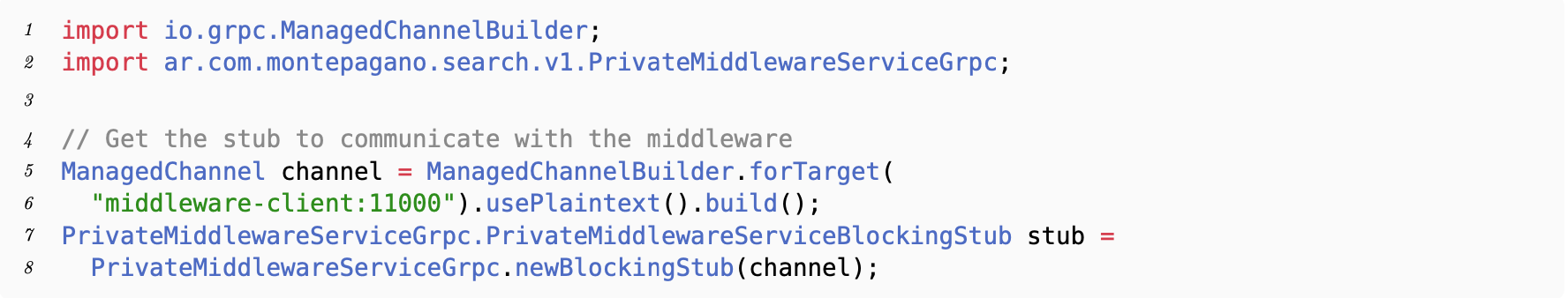}
\end{figure}

Relying on the snippet above, the next one shows how the client
 registers the communication channel to the middleware:
\begin{figure}[h]\centering
 \includegraphics[width=\textwidth]{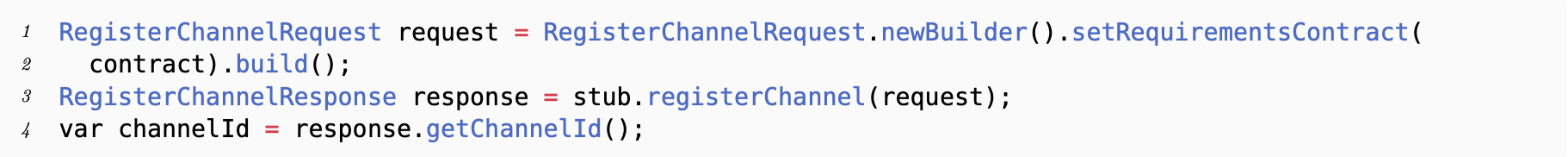}
\end{figure}
The object \texttt{contract}, passed to the function
\texttt{SetRequirementsContract}, is an instance of the Java class
\texttt{GlobalContract} (automatically generated from Protocol Buffer
type), built from the specification in \cref{fig:channel-spec}.

The snippet shown below, exhibits the client application invoking the middleware's operation \texttt{AppSendRequest} for sending a message \texttt{PurchaseRequest} (see line $3$ of \cref{fig:channel-spec}), to participant \texttt{Srv}, over the channel identified by \texttt{channelId} (the channel identifier is received at the moment of the registration of communication channel, see line $6$ of channel registration snippet).
\begin{figure}[h]\centering
\includegraphics[width=.98\textwidth]{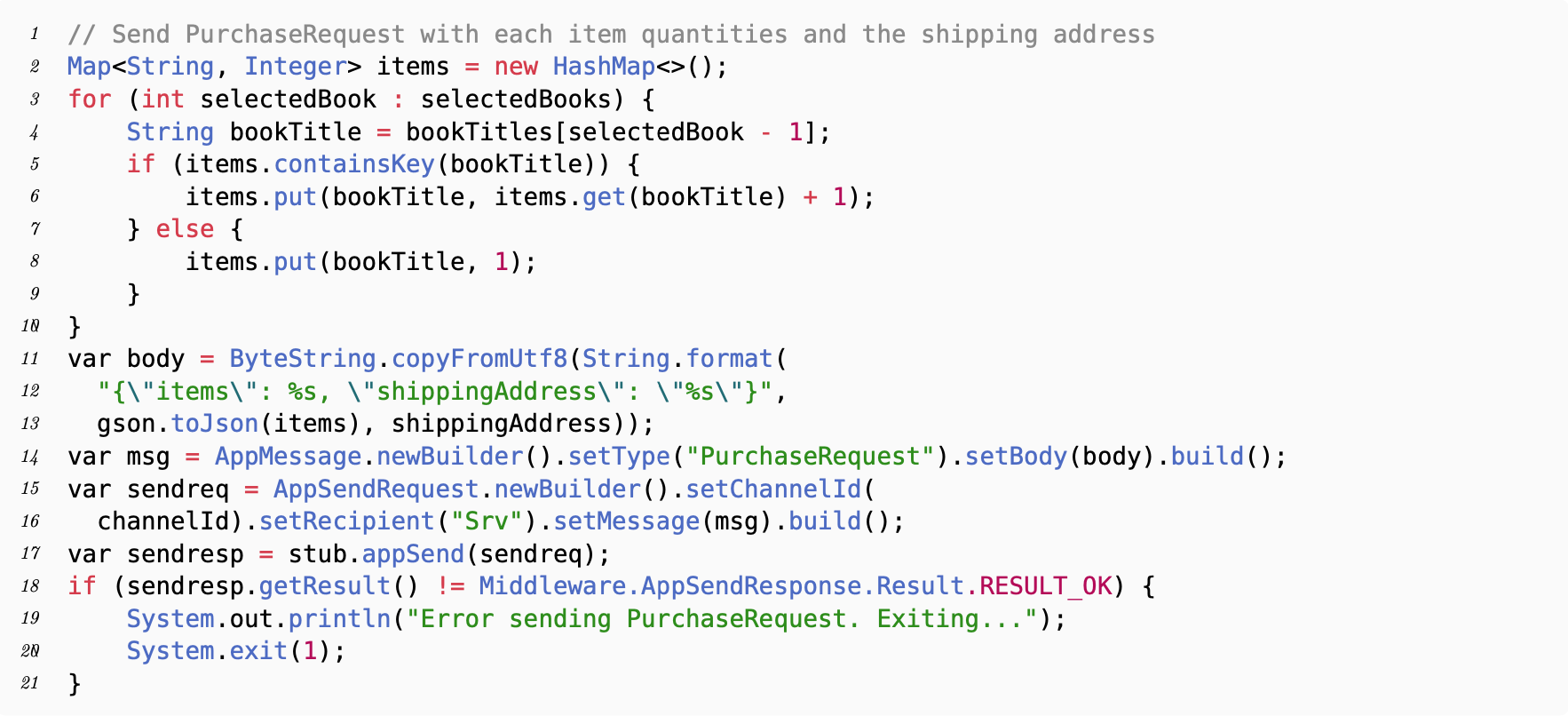}
\end{figure}
Analogously, the snippet below shows the client application invoking the middleware's operation \texttt{AppRecvRequest} for receiving a message \texttt{TotalAmount} (see line $4$ of \cref{fig:channel-spec}), from participant \texttt{Srv}, also over the channel identified by \texttt{channelId}.
\begin{figure}[h]\centering
\includegraphics[width=.98\textwidth]{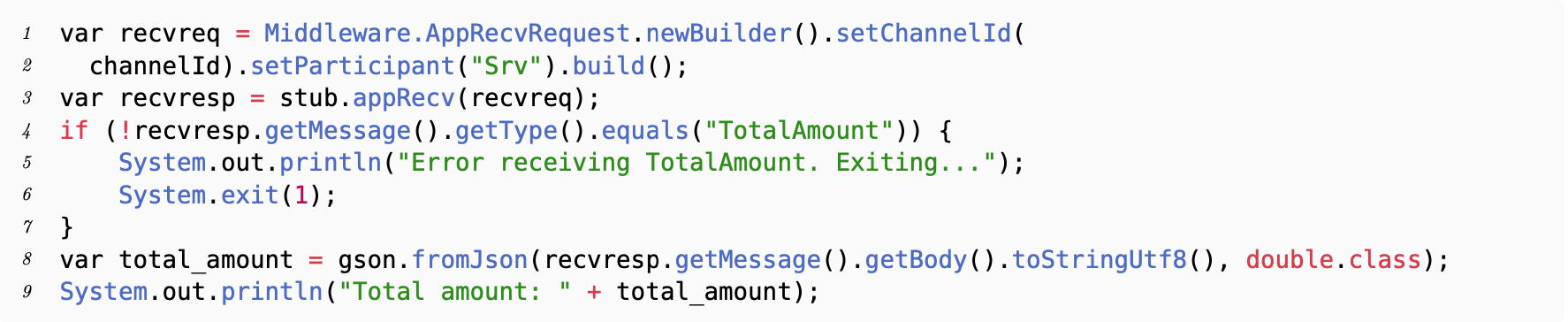}
\end{figure}

\subsection{Required services}
The seller's server application was developed in
Python; the snippet shown below exhibits the procedure for
registering the backend server through a gRPC channel received as
parameter.
\begin{figure}[t],
\centering
 \includegraphics[width=\textwidth]{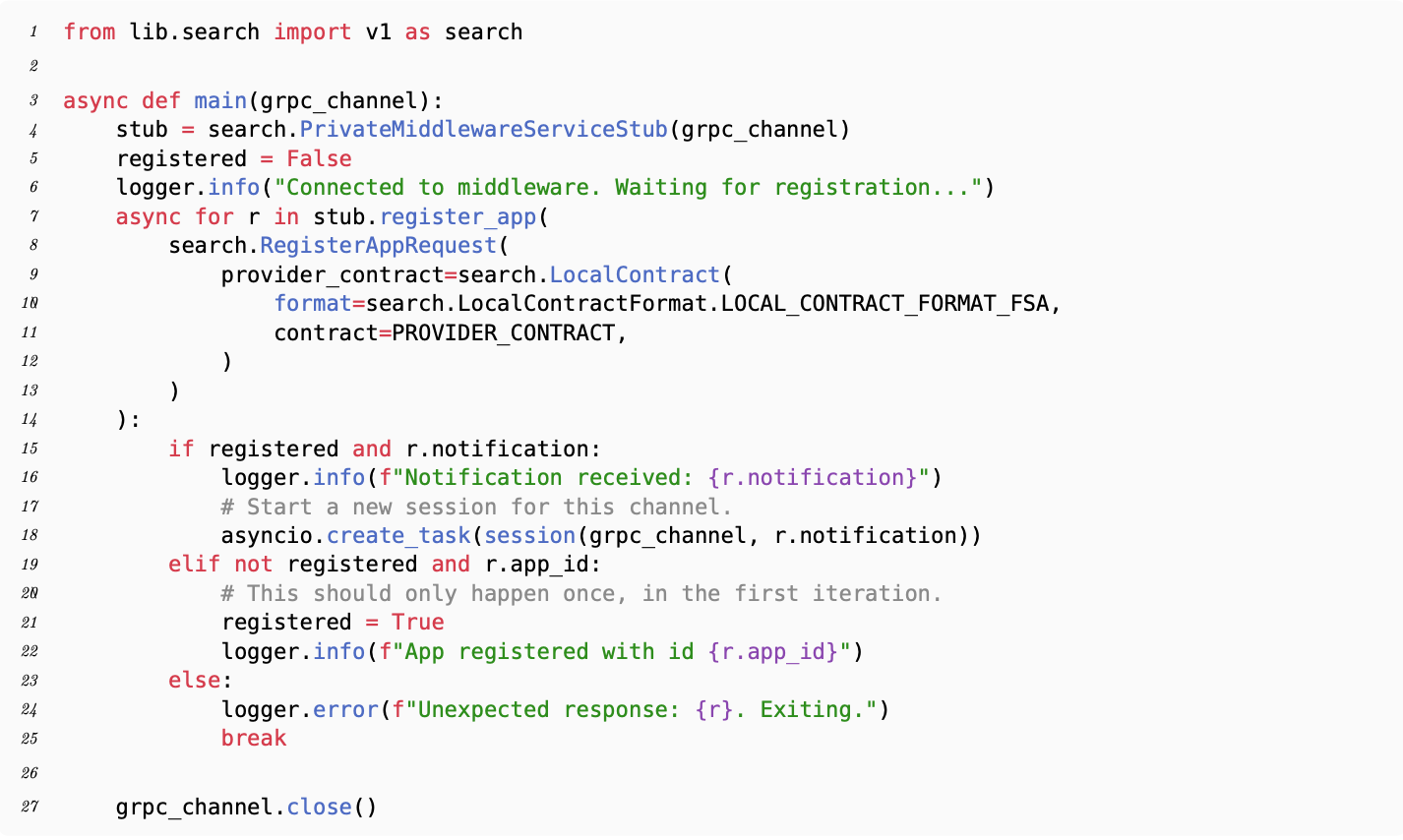}
\end{figure}

The process starts on line $4$ by opening a communication channel
through the middleware's private interface, by resorting the Python
stubs generated by Protocol Buffer.
Then, the middleware's operation \texttt{RegisterAppRequest} is
invoked with a local contract (i.e., a CFSM described as a single
finite state machine like the ones shown \cref{fig:channel-spec}). The
result of the process depends on whether the service has been already
registered or not; if the service has been properly registered by the
broker, the latter will return an application identifier for the
middleware to be able to refer to the service in its local host. Line
$18$ shows the asynchronous
invocation to the function \texttt{session} which implements the
server.

We omit the details of the Go implementation of the
payment service because they are analogous to the
seller implementation.

 \section{Conclusions and future work}
\label{sec:conclusions}

In this work we combined well established languages and tools from the
fields of service-oriented architectures, language semantics, and
behavioural types to develop \SEArch.
The execution infrastructure of \SEArch hinges as ARNs and has a
complete operational semantics that enables analysis based on LTL
formulae (see~\cite[Sec~4]{vissani:wadt14-f}).
Behavioural types, more specifically CFSMs, were used as
interoperability contracts that can be automatically analysed, thus
providing the means for checking service compliance with respect to a
requirement contract.
This last feature provides an answer to the problem of automatic
service discovery at runtime.
A careful selection of tools allowed us to implement a middleware and
a service broker that jointly provide transparent creation and
deletion of communication sessions, according to high-level
behavioural contracts.
This yields a general dynamic reconfiguration mechanism for this type
of service-based software artefacts.

The infrastructure of \SEArch is in a functional prototype stage and,
consequently, the implementation left space for many extensions.
Some of them are related to different aspects of scalability, for
example, the current implementation features a service repository
coupled to the implementation of the service broker.
An alternative, and more scalable, design might implement the
repository as a separate agent, allowing horizontal scaling of the
role and separating the registration process from the broker service.
This might also enable the broker to access multiple
repositories.
Another hurdle for scalability resides the centralised implementation
of the compliance check in the service broker; separating the analysis
of contracts as a service used by the broker might allow
implementation over clusters of computers that might even perform
off-line checks for precomputing the content of the cache we proposed,
and implemented, to make the brokering more efficient.

From~\cref{fig:execution} it is easy to observe that choosing a service
relies on abstract notions of provision and requirement contract. As
we mentioned in~\cref{sec:introduction}, CFSMs has been extended with
both, functional information and quality-of-service non-functional
information making QoS-enriched Data-aware CFSMs a type of contract
fit for describing many aspects relevant for both, service compliance
and service selection. We plan to implement both extensions, and their
associated verification algorithms, within
\chorgram~\cite{lange:gay17,chorgram}.

The compliance mechanism featured by \SEArch
relies on bisimilarity of CFSMs.
An interesting line of work is to embed in \SEArch other compliance
mechanisms based on different types of contracts, and their associated
tools.
Some options are tools like \toolid{CAT}~\cite{bdft16} which is based
on \emph{contract automata}~\cite{bt22,btp20,bt23} or
contract-oriented middlewares like the one in~\cite{bcmpp17,bcmpp15}
which supports timed behavioural types or the one in~\cite{abmtz17}.

Recently tools for inferring behavioural specifications from code have
been proposed.
For instance, \toolid{KmcLib}~\cite{imai:tacas22} extracts CFSMs from
Ocaml code, the tool in~\cite{vr17} infers behavioural types from Java
code, and \toolid{ChorEr}~\cite{gen23} extracts choreography
automata~\cite{blt20} from Erlang code,
\toolid{Contractor}\footnote{Available at
  \url{http://lafhis.dc.uba.ar/dependex/contractor/Welcome.html}.} is a
tool that extracts automata models from C programs.
Composing these tools with \SEArch opens the possibility of smoothly
integrate services to our infrastructure.

Lastly, the correct execution of software system in \SEArch requires
that the implementation of service providers honours the contract they
expose.
This may not hold for incorrect implementations or in adversarial
settings where malicious providers could be present.
In this work we adopted an approach in which the act of invoking the
middleware's operation \texttt{RegisterAppRequest}, and consequently
triggering the invocation of the service broker's operation
\texttt{RegisterProviderRequest} makes the provider fully responsible
for any inconsistency that might occur during execution.
However, \SEArch does not provide any mechanism for assigning the
blame.
Detecting these types of violations can be attained statically (e.g.,
with approaches like the one in~\cite{bstz16,bstz13} or using
behavioural contracts to synthesise a program skeletons which, after
the implementation, could be statically analysed by tools akin to
\toolid{Dafny}~\cite{leino:lpar10}.
However, runtime verification is required when code cannot be analysed
(e.g., third-party components).
In particular, one can use monitors capable of auditing the
communication session, according to the global contract specifying the
communication channel.
Under this view, it is paramount to provide analysis tools for the
development of services in order to ensure compliance between the
implementation and the provision contract.
Extending \SEArch for runtime verification is scope for future work.

\bibliographystyle{splncs}

\end{document}
